\documentstyle[12pt,epsf]{article}
\def\lQ{\Lambda_{\rm QCD}}

\newcommand{\be}{\begin{equation}}
\newcommand{\ee}{\end{equation}}
\newcommand{\bea}{\begin{eqnarray}}
\newcommand{\eea}{\end{eqnarray}}
\def\al{\alpha}
\def\als{\alpha_{\rm s}}
\def\siml{{\
\lower-1.2pt\vbox{\hbox{\rlap{$<$}\lower6pt\vbox{\hbox{$\sim$}}}}\ }}

\newcommand{\MS}{\overline{\rm MS}}
\newcommand{\RS}{\rm RS}

\newcommand{\OS}{\rm OS}

\begin{document}
\begin{titlepage}
\begin{flushright}
\tt{UB-ECM-PF 02/11}
\end{flushright}

\vspace{1cm}
\begin{center}
\begin{Large}
{\bf The static potential: lattice versus perturbation theory in a
renormalon-based approach}\\[2cm]
\end{Large} 
{\large Antonio Pineda}\footnote{pineda@ecm.ub.es}\\
{\it Dept. d'Estructura i Constituents de la Mat\`eria and IFAE,
  U. Barcelona \\ Diagonal 647, E-08028 Barcelona, Catalonia, Spain
        \\}
\end{center}

\vspace{1cm}

\begin{abstract}
We compare, for the static potential and at short distances,
perturbation theory with the results of lattice simulations. We show
that a renormalon-dominance picture explains why in the literature
sometimes agreement, and another disagreement, is found between
lattice simulations and perturbation theory depending on the different
implementations of the latter. We also show that, within a
renormalon-based scheme, perturbation theory agrees with lattice
simulations.
\vspace{5mm}  
\end{abstract}

\end{titlepage}
\vfill
\setcounter{footnote}{0} 
\vspace{1cm}

\section{Introduction}

The static potential is the object more accurately studied by
(quenched) lattice simulations. This is due to its relevance in order
to understand the dynamics of QCD. On the one hand, it is a necessary
ingredient in a Schr\"odinger-like formulation of the Heavy
Quarkonium.  On the other hand, a linear growing behavior at long
distance is signaled as a proof of confinement. Moreover, throughout
the last years, lattice simulations \cite{latticeshort1,NS,latticeshort2}
have improved their predictions at short distances allowing, for the
first time ever, the comparison between perturbation theory and
lattice simulations. Therefore, the static potential provides a unique
place where to test lattice and/or perturbation theory (depending on
the view of the reader), as well as an ideal place where to study the
interplay between perturbative and non-perturbative physics. This is
even more so since the accuracy of the perturbative prediction of the
static potential has also improved significantly recently
\cite{FSP,short,KP1,RG}.

\medskip

Let us first review the status of the art nowadays.  The prediction
for the perturbative static potential at two loops \cite{FSP}
indicated the failure (non-convergence) of perturbation theory at
amazingly short distances. This failure of perturbation theory is not
solved by the inclusion of the leading logs at three loops computed in
\cite{short,KP1} nor by performing a renormalization group (RG)
improvement of the static potential at next-to-next-to-leading log
(NNLL) \cite{RG}.\footnote{ There also exists a computation of the
running of the Coulomb potential in vNRQCD \cite{HMS} with the same
precision that disagrees with the one obtained in pNRQCD \cite{RG}. At
this respect, we would like to report on a recent computation
\cite{PS} of the 4-loop double log term of the Coulomb potential
proportional to $C_A^3\beta_0$ that agrees with the pNRQCD result and
disagrees with the vNRQCD one.}

On the other hand, it was soon realized that the static potential
suffered of renormalons \cite{Aglietti} and that the leading one (of
$O(\lQ)$) cancelled with the leading renormalon of twice the pole mass
\cite{thesis}. Nevertheless, the prediction of perturbation theory for
the slope of the static potential should not suffer, in principle,
from this renormalon and could be compared with lattice simulations,
which, indeed, only predict the potential up to a 
constant. This comparison was performed in Ref. \cite{Bali}, where
they compared the RG improved predictions
(without including ultrasoft log resummation) of perturbation theory
up to next-to-next-to-leading order (NNLO) with lattice
simulations. They indeed found that the discrepancies with lattice
simulations and the lack of convergence of the perturbative series
remained. They also found that the difference between perturbation
theory and lattice could be parameterized by a linear potential in a
certain range. Therefore, it seemed to support the claims of some
groups \cite{GPZ,S} of the possible existence of a linear potential at
short distances. Such claims contradict the predictions of the
operator product expansion (OPE), which state that the leading
non-perturbative corrections are quadratic in distance at short
distances\footnote{This contradiction is indeed so only if one 
believes that perturbation theory can be applied for the shortest 
distances available in lattice simulations $\sim 1-4$ GeV. We 
thank V.I. Zakharov for stressing this point to us.}.

On the other hand, driven by the analysis of Refs. \cite{BSV}, it has 
been argued \cite{Sumino} that, within a renormalon based picture,
perturbation theory agrees with the phenomenological potentials aimed
to describe Heavy Quarkonium. For his study, he used the static
version of the 1S mass and the upsilon expansion \cite{HLM}, which
cancels the leading renormalon and relate the 1S mass to the $\MS$
mass. Moreover, In Ref.  \cite{DESY} (see also \cite{Sumino,BB}), it
has been shown that perturbation theory can indeed reproduce the slope
of the static potential given by lattice simulations at short
distances by using the force instead of the potential as the basic
tool without the need to talk about renormalons.

\medskip

In this paper, we would like to try to clarify further the above
issues and support the renormalon dominance picture by comparing the
potential computed in lattice with the RS static potential (a
renormalon free definition of the potential obtained in
Ref. \cite{RS}).

\section{On-shell scheme}

The energy of two static sources in a singlet configuration reads (in
the on-shell (OS) scheme)
\be
E(r)=2m_{\OS}+\lim_{T\to\infty}{i\over T} \ln \langle W_\Box \rangle
\,.
\ee
In the situation $\lQ \ll 1/r$, it can be computed order by order in
$\als(\nu)$ ($\nu \sim 1/r$) and in the multipole expansion (see
\cite{pNRQCD,RG}) 
\be
E(r)=2m_{\OS}+V_{s,\OS}(r,\nu_{\rm US})+\delta E_{{\rm US}}(r,\nu_{\rm US})
\,.
\ee
$m_{\OS}$ is the pole mass. The static potential reads
($\als\equiv \als(\nu)$) 
\be
\label{potnu}
V_{s,\OS}\equiv V_{s,\OS}(r,\nu_{\rm US})=\sum_{n=0}^\infty V_{n} \als^{n+1},
\ee
where $V_n \equiv V_n(r,\nu,\nu_{\rm US})$. The first three
coefficients $V_0$, $V_1$ and $V_2$ are known \cite{FSP}. The
log-dependence on $\nu$ of $V_3$ can also be obtained by using the
$\nu$-independence of $V_{s,\OS}$. The log-dependence on $\nu_{\rm
US}$ of $V_3$ is also known \cite{short}. Therefore, the only unknown
piece of $V_3$ is a $\nu/\nu_{\rm US}$-independent constant. Its size
has been estimated in Ref. \cite{RS} assuming renormalon
dominance. For $n_f=0$, it reads
\be
\label{V3}
V_{3}(r,1/r,1/r)=1/r \times (-76.1075) 
\,.
\ee
Leaving aside the leading-renormalon independent contributions to
$V_{3}$, this number suffers from two sources of errors (see Ref. 
\cite{RS} for details). One is the
error in the evaluation of the normalization factor of the renormalon,
$N_m$, which we will discuss in the next section. The other error is
due to $O(1/n)$ corrections ($n=3$ in this case), which appear to be
negligible. The relative $O(1/n)$ corrections are -0.0264977, and the
$O(1/n^2)$ corrections are -0.00544012. Therefore, we expect the
unknown $O(1/n^3)$ effects to be very small.  We will use Eq. (\ref{V3}) 
in the following for our estimates of $V_3$.

\medskip

It is quite remarkable that the numbers obtained in Ref. \cite{RS} for
$V_3$ are quite close to the ones obtained in Ref. \cite{CE} using
Pad\'e-approximants methods. This is even so if we do the Fourier
transform back to momentum space. We display in Table \ref{tablec0}
the numbers one obtains from Ref. \cite{RS} for the coefficient $c_0$
(the portion of the three-loop momentum space coefficient of the
static potential independent of scale logarithms as defined in
Ref. \cite{CE}) for different number of massless flavours. We can see
that the difference with Chishtie and Elias results (see table 1 in
Ref. \cite{CE}) is always of the order of 20-30 and roughly
independent of the number of flavours. Therefore for $n_f \rightarrow
0$, where the absolute value of $c_0$ increases, the bulk of $c_0$
agrees with Chishtie and Elias prediction.

\begin{table}[h]
\addtolength{\arraycolsep}{0.2cm}
$$
\begin{array}{|l||c|c|c|c|c|c|c|}
\hline
n_f  & 0 & 1 & 2 & 3 & 4  & 5 & 6
\\ \hline
c_0 & 292 & 227 & 168  & 116 & 72 & 37 & 12 \\ \hline
\end{array}
$$
\caption{{\it Values of $c_0$, as defined in Ref. \cite{CE}, according
to the results of Ref. \cite{RS}.}} 
\label{tablec0}
\end{table}

\medskip

The static potential is $\nu$-independent. By setting $\nu=1/r$, we
could effectively resum the $\ln{\nu r}$ terms. The RG-improved
expressions would read
\be
\label{potRG}
V_{s,\OS}=\sum_{n=0}^\infty V_{n}(r,1/r,\nu_{\rm US}) \als(1/r)^{n+1},
\ee
and we would have expressions for $n=0,1,2,3$. In the above expression
we have considered $\ln r\nu_{\rm US} \sim 1$. Since $\ln r\nu_{\rm
US} \gg 1$ for some range of the parameters, one could be in the
situation where one also has to resum ultrasoft logs as it has been
done in Ref. \cite{RG}. Nevertheless, explicit calculations show that,
at least for the range of parameters studied in this paper, higher
order ultrasoft logs are subleading even if sizable. However, for
definiteness, we will work with the resummed expression (the physical
picture does not change anyhow), which we will add to
$V_{3}(r,1/r,1/r)$:
\be
V_{3}(r,1/r,\nu_{\rm US}) \equiv V_{3}(r,1/r,1/r)
+
{C_A^3\over
  6\beta_0}\als^3(r^{-1}) \log\left(
\alpha_{s}(r^{-1})\over \alpha_{s}(\nu_{us}) \right)
\,.
\ee
In principle, it could be considered to be more consistent to add the
resummed expression for the ultrasoft logs to $V_2$. Nevertheless, we
will add them to $V_3$ since it is the counting consistent when the
leading single log gives the bulk of the correction (moreover, we will
see that, even after the leading-renormalon cancellation is achieved,
its effect is small\footnote{However, it remains to be seen whether,
once $V_3$ is exactly computed and the $O(r^2)$ renormalon were also
subtracted, the ultrasoft logs would be a subleading effect compared
with finite pieces.} if compared with the typical estimate of the
terms of $V_3$).

\medskip

The leading $O(r^2)$ non-vanishing contribution to $\delta E_{{\rm
US}}(r,\nu_{\rm US})$ reads (we actually write the Euclidean
expression)
\begin{equation}
\delta E_{\rm US}(r,\nu_{us}) \simeq  {T_F \over 3 N_c} {\bf r}^2 \int_0^\infty \!\! dt 
 e^{-t(V_{o,\OS}-V_{s,\OS})} \langle g{\bf E}^a(t) 
\phi(t,0)^{\rm adj}_{ab} g{\bf E}^b(0) \rangle(\nu_{us}).
\label{energyUS}
\end{equation}
At the order of interest $V_{o,\OS} \equiv \displaystyle{{1 \over
2N_c}{\als \over r}}$ and $V_{s,\OS}$ is approximated to its leading
order value: $-C_F \displaystyle{{\als \over r}}$. This contribution
cancels the leading $\nu_{\rm US}$ scale dependence of $V_{s,\OS}$.
For the scales we will study in this paper (those accessible by
lattice simulations), we will consider $\als/r$ to be a
non-perturbative scale. Therefore, Eq. (\ref{energyUS}) can not be
computed perturbatively and it will not be considered in this and the
following section, where we aim to a pure perturbative prediction (it
will be considered, however, in sec.  \ref{quadpot}). Thus, our
results will depend on $\nu_{us}$, for which we will take
$\nu_{us}=2.5 \,r_0^{-1}$. We will set $n_f=0$ and work in lattice
units: $r_0^{-1}$ (with $r_0^{-1} \simeq 400\, {\rm MeV}$ according to
lattice simulations), $\Lambda_{\MS} = 0.602(48)\,r_0^{-1}$
\cite{Lambda}, since our aim will be to compare with lattice
simulations.

\medskip

\begin{figure}[h]
\hspace{-0.1in}
\epsfxsize=4.8in
\centerline{
\put(-37,100){$r_0V_{\OS}(r)$}
\epsffile{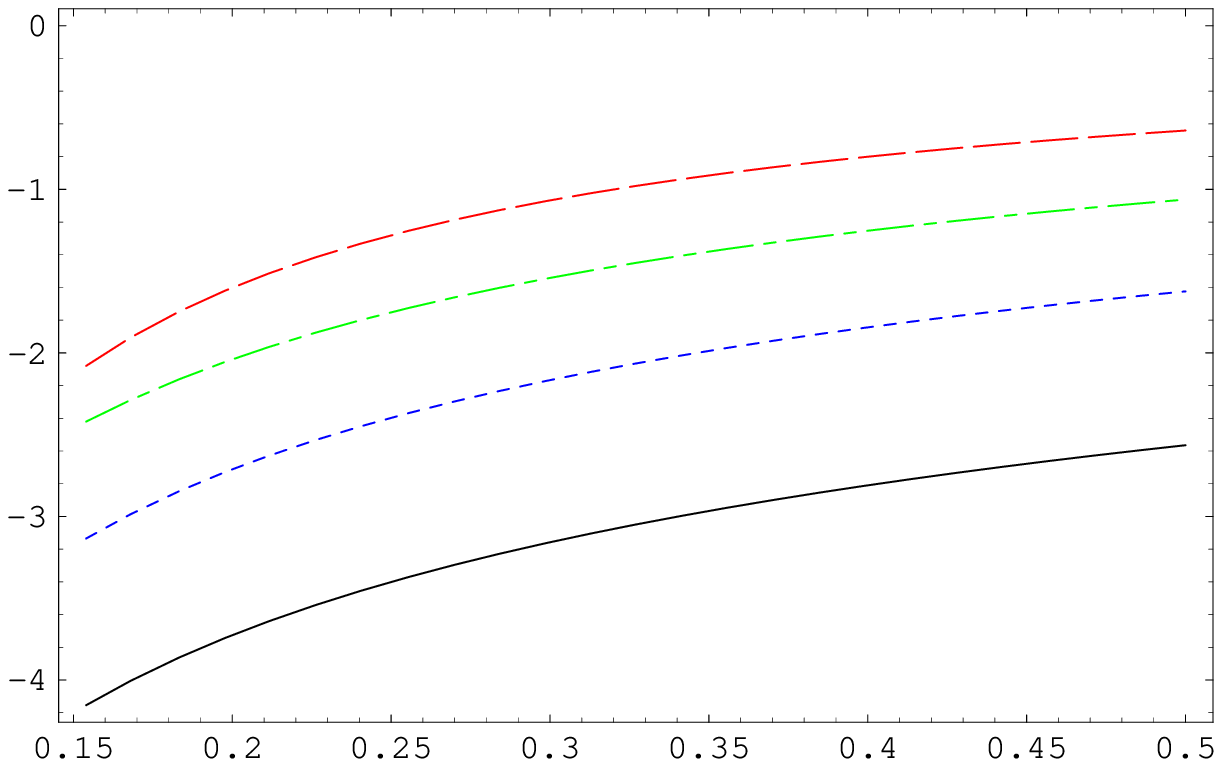}
\put(15,1){$r/r_0$}
}
\caption {{\it Plot of $r_0V_{\OS}(r)$ at tree (dashed line), one-loop
(dash-dotted line), two-loops (dotted line) and three loops (estimate)
plus the leading single ultrasoft log (solid line). For the scale of
$\als(\nu)$, we set $\nu=1/0.15399\,r_0^{-1}$. $\nu_{us}=2.5\,r_0^{-1}$.}}
\label{potOSnu}
\end{figure}
\begin{figure}[h]
\hspace{-0.1in}
\epsfxsize=4.8in
\centerline{
\put(-37,100){$r_0V_{\OS}(r)$}
\epsffile{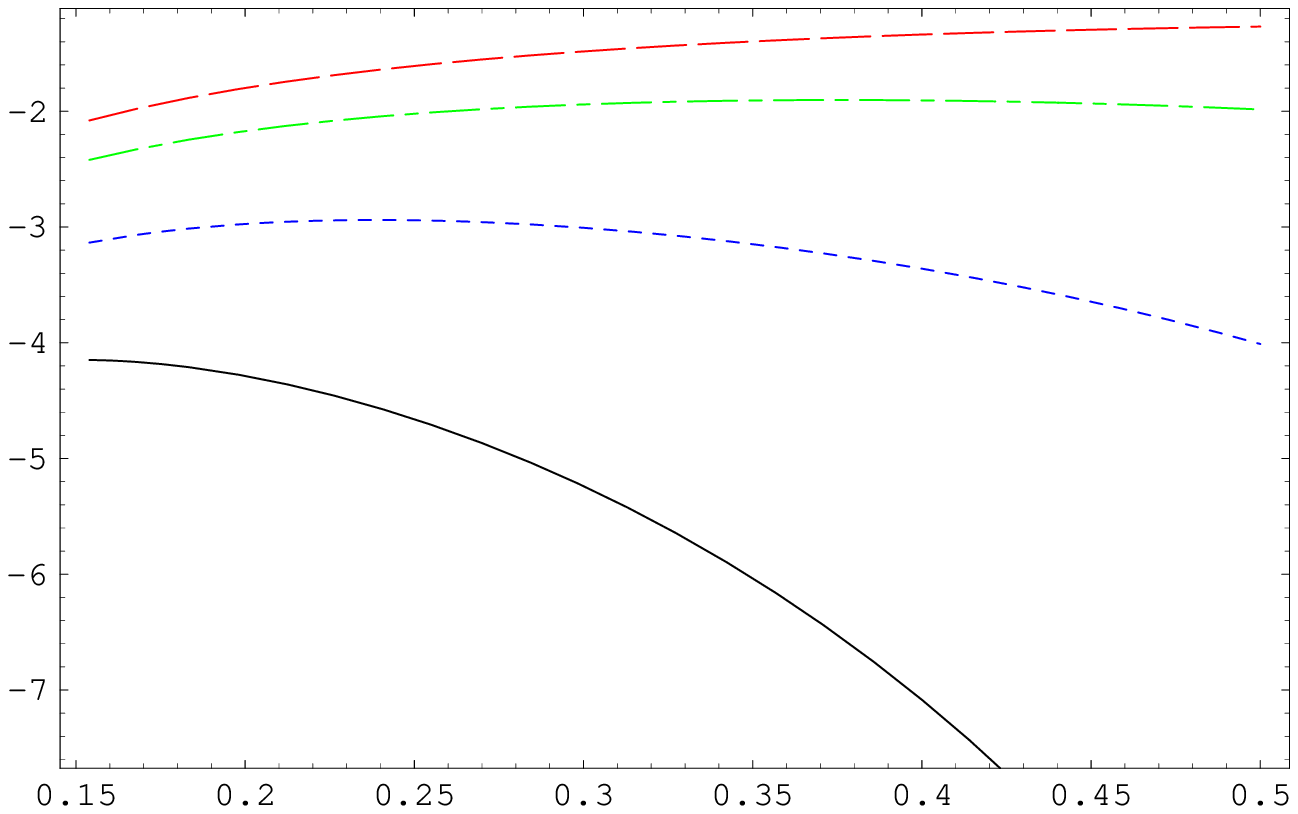}
\put(15,1){$r/r_0$}}
\caption {{\it Plot of $r_0V_{\OS}(r)$ at tree (dashed line), one-loop
(dash-dotted line), two-loops (dotted line) and three loops (estimate)
plus the RG expression for the ultrasoft logs (solid line). For the
scale of $\als(\nu)$, we set $\nu=1/r$. $\nu_{us}=2.5\,r_0^{-1}$.}}
\label{potOSr}
\end{figure}

Now, we want to check the convergence of the series within
perturbation theory. We first consider the perturbative static
potential without trying to resum $\ln r\nu$ logs, ie. for a fixed
scale $\nu$ for $\als(\nu)$ (Eq. (\ref{potnu})). We show our results
in Fig. \ref{potOSnu}, where we have chosen $\nu=1/0.15399\,r_0^{-1}$. We see
that it does not converge to a value. We next try with $\ln r\nu$
resummation (Eq. (\ref{potRG})), since these logs could be large. We
show our results in Fig. \ref{potOSr}. We find the same problem and
moreover the slope appears to be different in both cases. The
situation appears to be puzzling.

\medskip

Let us now compare with lattice simulations. These predict the energy
of an static quark-antiquark system up to an $r$-independent
constant\footnote{Therefore, due to the unknown constant it can not
be related with the underlying theory (QCD) and, for instance, obtain
the mass.}:
\be
E_{\rm latt.}(r)=k_{\rm latt.}+\lim_{T\to\infty}{i\over T} \ln \langle
W_\Box \rangle .
\ee
Therefore, the slope of the potential is a physical prediction of the
lattice static potential and does not suffer from the leading
renormalon. The constant is usually fixed in a way such that
\be
E_{\rm latt.}(r_0)=0
\,,
\ee
except in Ref. \cite{NS} where $E_{\rm latt.}(r_c)=0$ is used 
(for the definition of $r_c$ see \cite{NS}). 

In order to compare the perturbative predictions with lattice
simulations, we add a constant to the perturbative potential to fix
both potentials to be equal at some scale $r'$:
\be
E_{\rm latt.}(r')=k_{\rm per.}(r')+V_{s,OS}(r')
\,.
\ee
The value of this constant varies according to the order the perturbative
computation has been done, ie.  
\be
k_{\rm pert.}(r')=k_0(r')+k_1\als(r')+k_2\als^2(r')+\cdots 
\,.
\ee 

We will consider the lattice data of Refs. \cite{latticeshort1,NS}.
They both appear to be perfectly compatible with each other for finite
lattice spacings. In any case, the lattice data of Ref. \cite{NS},
being more recent, appears to have smaller errors with a very small
spread of the points around the fitted curve.  Here, in order to get
the highest possible precision, we will use the lattice data of
Ref. \cite{NS} for which the continuum limit has been reached. For
comparison with the lattice points at finite lattice spacing of
Ref. \cite{latticeshort1} and for a somewhat a larger range, we refer
to the 1st version of this paper sent to the web. The physical outcome
is, in any case, the same.

\begin{figure}[h]
\hspace{-0.1in}
\epsfxsize=5.2in
\centerline{
\put(50,190){$r_0(V_{\OS}(r)-V_{\OS}(r')+E_{latt.}(r'))$}
\epsffile{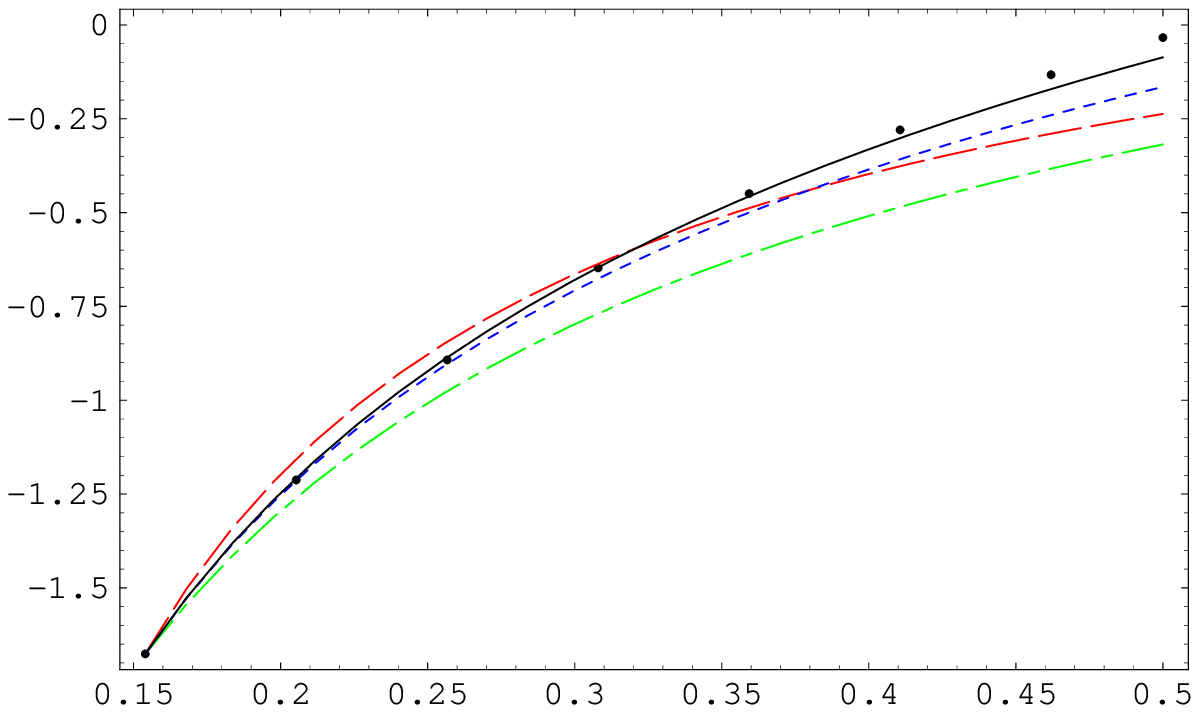}
\put(15,1){$r/r_0$}}
\caption {{\it Plot of $r_0(V_{\OS}(r)-V_{\OS}(r')+E_{latt.}(r'))$ versus
$r$ at tree (dashed line), one-loop (dash-dotted line), two-loops
(dotted line) and three loops (estimate) plus the leading single
ultrasoft log (solid line) compared with the lattice simulations
\cite{NS} $E_{latt.}(r)$. For the scale of $\als(\nu)$, we
set $\nu=1/0.15399\,r_0^{-1}$. $\nu_{us}=2.5\,r_0^{-1}$ and
$r'=0.15399\,r_0$.}} 
\label{potOSlattnu}
\end{figure}

\begin{figure}[h]
\hspace{-0.1in}
\epsfxsize=5.2in
\centerline{
\put(50,200){$r_0(V_{\OS}(r)-V_{\OS}(r')+E_{latt.}(r'))$}
\epsffile{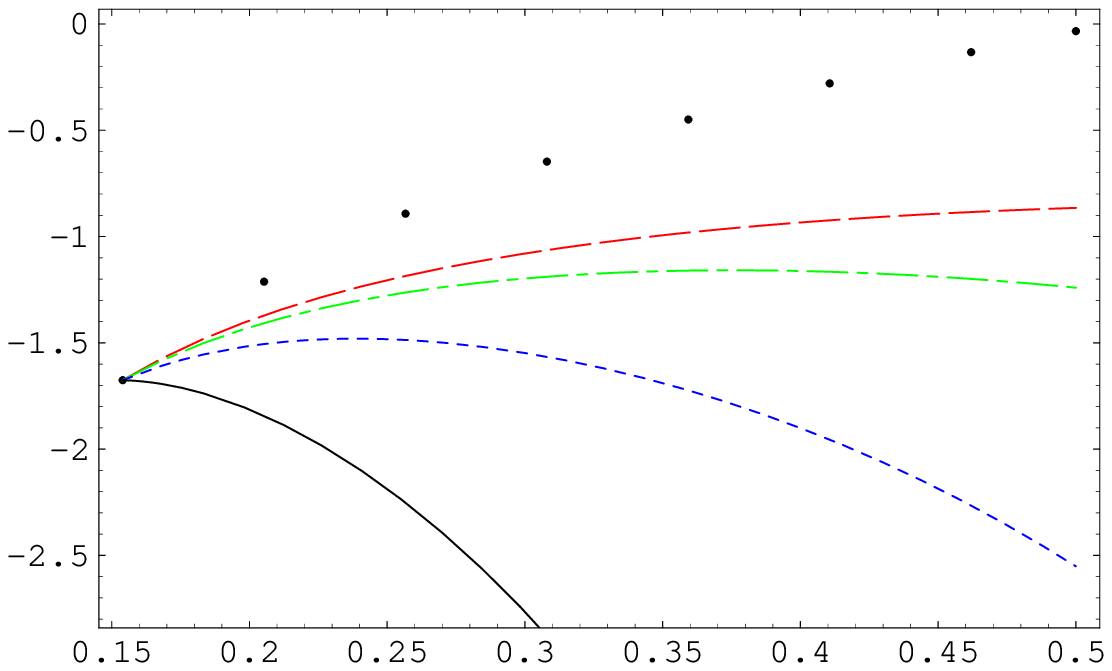}
\put(15,1){$r/r_0$}
}
\caption {{\it Plot of $r_0(V_{\OS}(r)-V_{\OS}(r')+E_{latt.}(r'))$ versus r
at tree (dashed line), one-loop (dash-dotted line), two-loops (dotted
line) and three loops (estimate) plus the RG expression for the
ultrasoft logs (solid line) compared with the lattice simulations
\cite{NS} $E_{latt.}(r)$. For the scale of $\als(\nu)$, we
set $\nu=1/r$. $\nu_{us}=2.5\,r_0^{-1}$ and $r'=0.15399\,r_0$.}}
\label{potOSlattr}
\end{figure}

The comparison with lattice is performed in Fig. \ref{potOSlattnu} for
the perturbative result without $\ln(r\nu)$ resummation ($\nu$
constant) and in Fig. \ref{potOSlattr} for the perturbative result
with $\ln(r\nu)$ resummation ($\nu=1/r$). As explained above, we have
added a constant to the perturbative potential (different at each
order) as to make it equal to the lattice potential at
$r´=0.15399\,r_0$. We see a completely different behavior in both
cases. Whereas for the latter, the agreement goes worse as we increase
the order of our calculation, for the former the agreement with
lattice improves\footnote{Usually, the comparison between lattice and
perturbation theory was performed using the expressions with
$\nu=1/r$, whereas the analysis with $\nu$ constant was largely
unnoticed.}. Indeed, it is quite remarkable that lattice simulations
are precise enough to see the log behavior of $\als$ predicted by
asymptotic freedom.

In the next section we will explain the above behavior (why sometimes
perturbation theory can describe lattice data and some others not)
within an controlled framework where the renormalon is taken into
account.

\section{RS scheme}

The RS scheme was defined in Ref. \cite{RS} aiming to eliminate the
renormalons of the matching coefficients appearing in Heavy Quarkonium
calculations. In particular, the leading renormalon of the heavy quark
mass and the static potential. We refer to Ref. \cite{RS} for
details. Here, we just write the relevant formulas needed for our
analysis.

Analogously to the OS scheme, the energy of two static sources in a
singlet configuration reads
\be
E_s(r)=2m_{\RS}(\nu_f)+\left(\lim_{T\to\infty}{i\over T} \ln \langle W_\Box
\rangle-2\delta m_{\RS}(\nu_f) \right)
\,,
\ee
where\footnote{Actually, we are going to use in this paper the RS'
scheme defined in Ref. \cite{RS} instead of the RS scheme, since we
believe it has a more physical interpretation. Nevertheless, we will
denote it in this paper as RS scheme in order to simplify the
notation. In any case, the physical picture does not change. The
results would also converge towards the result predicted by lattice
simulations but with a different pattern.}
\be
\delta m_{\RS}(\nu_f)=\sum_{n=1}^\infty \delta m_{\RS,n}\als^{n+1} =\sum_{n=1}^\infty  N_m\,\nu_f\,\left({\beta_0 \over
2\pi}\right
)^n \als^{n+1}(\nu_f)\,\sum_{k=0}^\infty c_k{\Gamma(n+1+b-k) \over
\Gamma(1+b-k)}
\,,
\ee
and $m_{\RS}\equiv m_{\OS}-\delta m_{\RS}$. If the beta function were
known to infinity order in perturbation theory, it would be possible
to obtain all the coefficients $b$ and $c_k$ \cite{Beneke2}. In
practice, only $b$ and $c_{0,1,2}$ are known (see
\cite{Beneke2,RS,renormalons}).  For $N_m$ only an approximate
calculation is possible using some ideas first developed in Ref.
\cite{Lee}. This computation has been done in Ref. \cite{RS}. The
result reads (for $n_f=0$)
\be
N_m=0.424413+0.174732+0.0228905=0.622036
\,,
\ee
where each term corresponds to a different power in $u$ (in the Borel
plane) of the calculation (see Ref. \cite{RS} for details). We see a
nice convergence. This number will be the one we will use in the
following. This number should be equal to $-2N_V$, where $N_V$ is the
normalization factor of the renormalon of the static potential. $N_V$
was also approximately computed in Ref. \cite{RS}. For $n_f=0$, it
reads
\be
N_V=-1.33333+0.499433-0.338437=-1.17234
\,.
\ee
In this case the convergence is not so good but we have an alternating
series. In any case, we see that both values appear to be quite
close:
\be
2{2N_m+N_V \over 2N_m-N_V} = 0.059
\,.
\ee
We take this as an approximate indication of the error in the
evaluation of $N_m$.

\medskip

In the situation $\lQ \ll 1/r$, $E_s(r)$ can be factorized in the
following way (with the accuracy we are working, the expression for
$\delta E_{s,{\rm US}}$ will be equal in the OS and RS scheme):
\be
E_s(r)=2m_{\RS}(\nu_f)+V_{s,\RS}(r;\nu_{us};\nu_f)+\delta E_{s,{\rm
US}}(r,\nu_{us}) 
\,.
\ee 
$V_{s,\RS}(r)\equiv V_{s,\RS}(r;\nu_{us};\nu_f)$ was defined in
Ref. \cite{RS} (note that we expand $V_s$ and $\delta m_{\RS}$ with
the same $\als$): \be V_{s,\RS}(r)=V_{s}+2\delta
m_{\RS}=\sum_{n=1}^\infty (V_{n}+2\delta m_{\RS,n})\als^n
=\sum_{n=0}^\infty V_{\RS,n}^{(0)} \als^{n+1}.
\ee
Due to the leading renormalon, the behavior of the perturbative
expansion at large orders of $V_{n}$ reads \cite{Beneke2,RS}
\be\label{generalV}
V_{n} \stackrel{n\rightarrow\infty}{=} N_V\,\nu\,\left({\beta_0 \over 
2\pi}\right)^n
 \,{\Gamma(n+1+b) \over
 \Gamma(1+b)}
\left(
1+\frac{b}{(n+b)}c_1+\frac{b(b-1)}{(n+b)(n+b-1)}c_2+ \cdots
\right)
.
\ee
Therefore, by adding $\delta m_{\RS}$ to $V_s$ we expect to cancel the
renormalon of the static potential. Indeed, what we want is to obtain
the renormalon cancellation order by order in perturbation
theory. This means to kill the $O(n!)$ behavior of the coefficient
multiplying the $O(\als^n)$ term of $V_{s}$ with the $O(n!)$ behavior
that appears in $\delta m_{\RS}$ such that $V_{\RS}$ enjoys nicer
convergence properties.  In principle, this cancellation should hold
for an arbitrary $\nu$ and, in particular, for $\nu=1/r$. This,
indeed, is the key observation of this paper and it will become
relevant later on in order to explain the results in the previous
section as well as in the literature.

\medskip

In principle, any $\nu_f$ should cancel the renormalon of the static
potential (we have the constraint, though, that $\nu_f$ should be
large enough to fulfill $\als(\nu_f) \ll 1$). If we re-expand $\delta
m_{\RS}(\nu_f)$ in terms of $\als(\nu)$, the renormalon is still kept
in $\delta m_{\RS}(\nu_f)$ but new terms are generated, which,
nevertheless, should not belong to the renormalon. In other words
\be
\label{analitic}
\delta m_{\RS}(\nu_f)=\delta m_{\RS}(\nu)+F(\nu;\nu_f)
,
\ee
where the Borel transform of $F(\nu;\nu_f)$ should be analytic at
$u=1/2$ in the Borel plane. This can be easily seen in the large
$\beta_0$ limit. In more formal terms what we have is 
$$
\Lambda_{\MS}=constant \,Im[\delta m_{\RS}(\nu_f)]
\,,
\qquad
\Lambda_{\MS}=constant \, Im[\delta m_{\RS}(\nu)]
$$
with the same constant by definition (indeed this is what defines the
renormalon). Therefore, the conclusion is that 
$$
Im[F(\nu;\nu_f)]=0
$$
to the order of interest, concluding that $F$ does not have the closest
singularity in the Borel plane. Note that in order to have the structure 
of Eq. (\ref{analitic}) is
necessary to have the {\it complete} renormalon contribution, ie. to
all orders in perturbation theory. If we have the renormalon to some
given order in perturbation theory, a change of the scale will produce
large (renormalon-related) terms that would not cancel since we do not
have the complete sum. Indeed, this is the explanation why in fixed
order perturbation theory one has to expand on $\als$ at the very same
scale for the renormalon and the OS result in order to achieve the
renormalon cancellation order by order in $\als$.

On the other hand, $\nu_f$ can not be arbitrarily large. Otherwise the
power counting is broken. This restricts $\nu_f \siml 1/r$ in order to
ensure the counting $V_{\RS} \sim O(\als/r)$. In our case, we will set
$\nu_f=2.5\,r_0^{-1} \sim 1\;{\rm GeV}$. In any case, when studying
the slope of the potential, the result should be independent on
$\nu_f$, as $\delta m_{\RS}(\nu_f)$ is just an r-independent
constant. At this respect, we should say that this is indeed so to a
large extent. If we take $\nu_f=5\,r_0^{-1} \sim 2\;{\rm GeV}$
agreement with lattice is also obtained\footnote{One could say that
since $\delta m_{\RS}$ starts at $O(\als^2)$ a larger value of $\nu_f$
could be used and yet not to break the power counting. Nevertheless,
we refrain from using $\nu_f \sim 5\, r_0^{-1}$ since we do not
understand the meaning of the results with values of $\nu_f$ larger
than $1/r$. In particular, this would have a difficult explanation
from an effective theory point of view, where $\nu_f$ is understood as
an cutoff of the effective theory, which has to be smaller than
$1/r$.}.  One could also think of setting $\nu_f=1/r$. This would mean
to subtract an $r$-dependent constant to the static potential. This is
not what it is done in lattice simulations since they are arbitrary up
to a $r$-independent constant. Therefore, this would not agree with
lattice simulations.

\medskip
\begin{figure}[h]
\hspace{-0.1in}
\epsfxsize=4.8in
\centerline{
\put(-37,100){$r_0V_{\RS}(r)$}
\epsffile{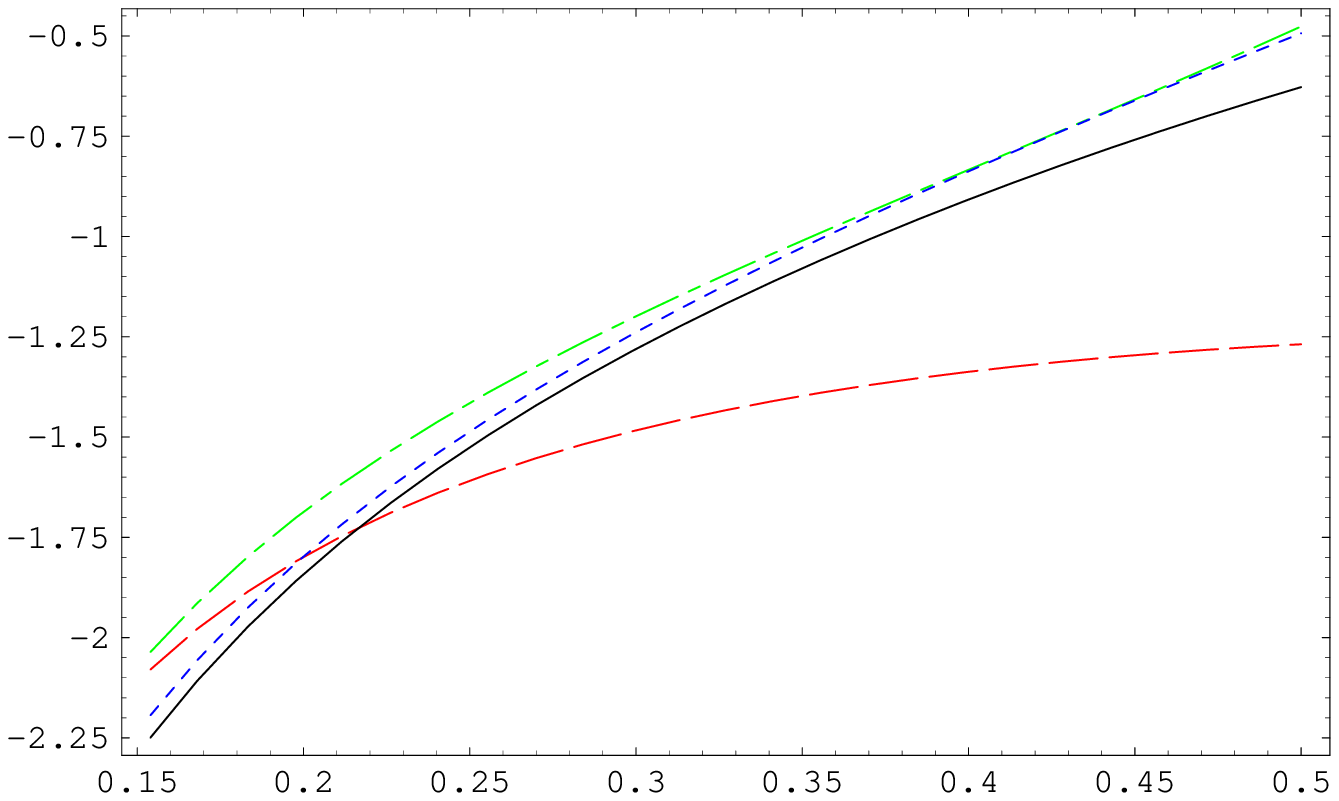}
\put(15,1){$r/r_0$}}
\caption {{\it Plot of $r_0V_{\RS}(r)$ at tree (dashed line), one-loop
(dash-dotted line), two-loops (dotted line) and three loops (estimate)
plus the RG expression for the ultrasoft logs (solid line). For the
scale of $\als(\nu)$, we set $\nu=1/r$. $\nu_{us}=2.5\,r_0^{-1}$ and
$\nu_f=2.5\,r_0^{-1}$.}}
\label{potRSr}
\end{figure}
\begin{figure}[h]
\hspace{-0.1in}
\epsfxsize=4.8in
\centerline{
\put(-37,100){$r_0V_{\RS}(r)$}
\epsffile{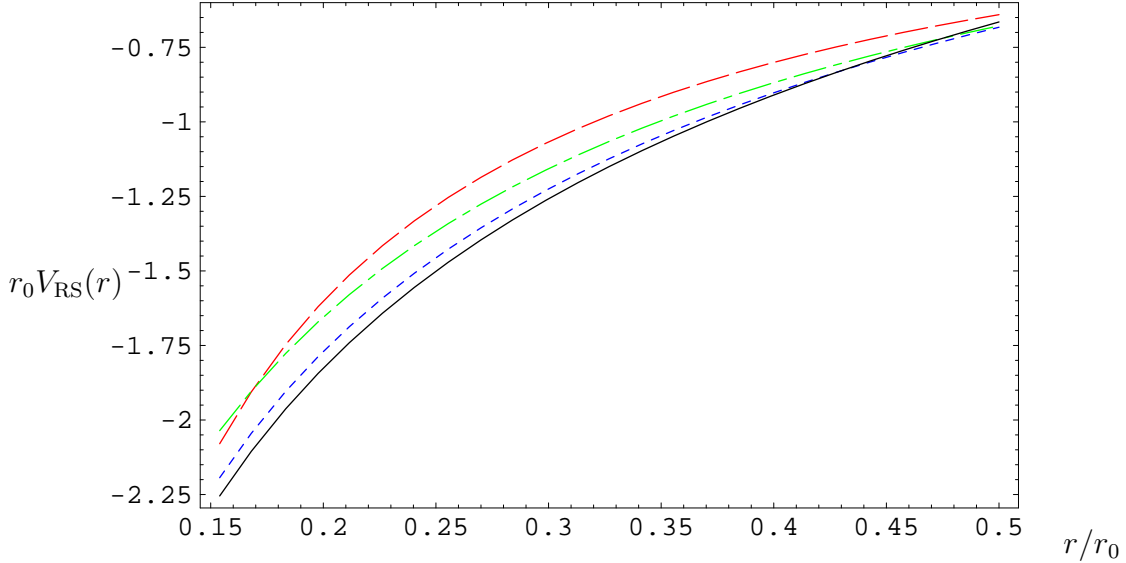}
\put(15,1){$r/r_0$}}
\caption {{\it Plot of $r_0V_{\RS}(r)$ at tree (dashed line), one-loop
(dash-dotted line), two-loops (dotted line) and three loops (estimate)
plus the leading single ultrasoft log (solid line). For the scale of
$\als(\nu)$, we set $\nu=1/0.15399\,r_0^{-1}$. $\nu_{us}=2.5\,r_0^{-1}$ and
$\nu_f=2.5\,r_0^{-1}$.}}
\label{potRSnu}
\end{figure}

In conclusion, by using $V_{\RS}$, we expect the bad perturbative
behavior due to the renormalon to disappear. Let us see that this is
indeed so. Working analogously to the OS scheme, we consider the RS
potential at different orders in perturbation theory with $\ln(r\nu)$
resummation or not. We display our results in Figs. \ref{potRSr} and
\ref{potRSnu}. In the first case, the $\ln(r\nu)$ resummation is
achieved by setting $\nu=1/r$ and the ultrasoft log resummation is
also included. In the second case, we set $\nu=1/0.15399\,r_0^{-1}$ and only
the leading single ultrasoft log is included. We see that now we do
not have the gap between different orders in the perturbative
expansion (or it is dramatically reduced). Moreover, we do not have a
(dramatically) different behavior of the slope of the potential;
$\ln(r\nu)$ resummation or not converges to the same value.

\medskip

\begin{figure}[h]
\hspace{-0.1in}
\epsfxsize=5in
\centerline{
\put(50,190){$r_0(V_{\RS}(r)-V_{\RS}(r')+E_{latt.}(r'))$}
\epsffile{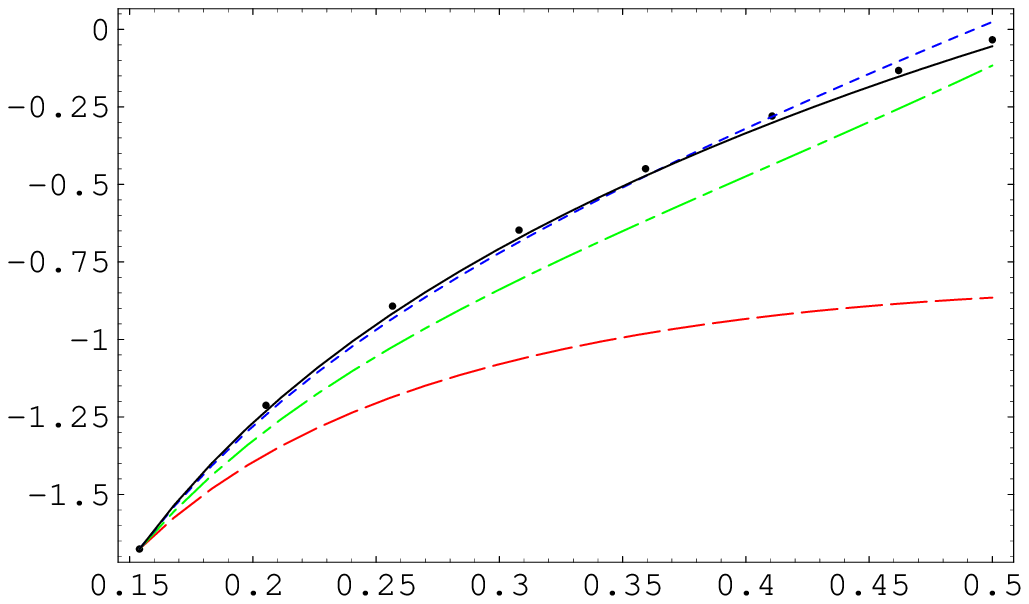}
\put(15,1){$r/r_0$}
}
\caption {{\it Plot of $r_0(V_{\RS}(r)-V_{\RS}(r')+E_{latt.}(r'))$ versus r
at tree (dashed line), one-loop (dash-dotted line), two-loops (dotted
line) and three loops (estimate) plus the RG expression for the
ultrasoft logs (solid line) compared with the lattice simulations
\cite{NS} $E_{latt.}(r)$. For the scale of $\als(\nu)$, we
set $\nu=1/r$. $\nu_f=\nu_{us}=2.5\,r_0^{-1}$ and $r'=0.15399\,r_0$.}}
\label{potRSlattr}
\end{figure}

We can now compare with lattice simulations: Figs. \ref{potOSlattnu}
and \ref{potRSlattr}. We work analogously to the OS scheme. Now, the
constant we have to add is approximately $\als$-independent, 
reflecting that we have accurately achieved the
renormalon cancellation. We see that our results are convergent to the
same potential, which, as we can see, corresponds to the lattice
potential.

The case without $\ln(r\nu)$ resummation is exactly equal to the OS
case, ie. to Fig. \ref{potOSlattnu}, since it is just equivalent to
add a r-independent constant at each order in $\als$. This explains,
within the renormalon dominance picture, the success of that specific
OS scheme calculation; it is just equivalent to an specific case of
the renormalon based calculation: $\nu=constant$.  Therefore,
perturbation theory without log resummation can explain lattice data
up to $r \sim 0.5\,r_0$ by subtracting a r-independent constant at
each order in perturbation theory.  This fact is understandable if the
factorization scale chosen is not very far of $1/r$ so that the RG
does not play a decisive role plus the fact that subtracting a
constant kills the renormalon.  However, now, we are not restricted to
take $\nu=constant$ but we can take $\nu=1/r$, which allows us to
resum the $\ln(r\nu)$ terms and yet cancel the renormalon as we can
nicely see in Fig. \ref{potRSlattr}.

\medskip

Within the renormalon dominance picture, we can also explain why
perturbation theory with $\ln(r\nu)$ resummation could not explain
lattice data (see Fig. \ref{potOSlattr}) even if subtracting a
r-independent constant at each order in perturbation theory. In
particular, this would explain the disagreement found in
Ref. \cite{Bali}.  The explanation of this fact is that $\ln(r\nu)$
resummation is equivalent to set $\als(r)$ as the expansion
parameter. Therefore, the expansion parameter is a function of $r$
and, consequently, the amount to be subtracted at each order in
$\als(r)$ in order to cancel the renormalon is also $r$-dependent. It
follows that the (perturbative) renormalon cancellation can not be
achieved by only subtracting a {\it $r$-independent} constant at each
order in $\als(r)$ as it was done in Fig. \ref{potOSlattr}.

We would like now to understand the agreement between lattice
\cite{DESY} (or potential models \cite{Sumino}) and perturbation
theory if the static potential is reconstructed from the force in the
following way:
\be
\label{force}
V(r)=\int_{r_0}^rdr'F(r')+V(r_0)=\int_{r_0}^rdr'{d V(r') \over dr'} +V(r_0)
\,.
\ee 
As far as we end up in the potential, we should have the same problems
(and solutions) than in the previous sections. The point is how
Eq. (\ref{force}) is implemented in practice.  The force also has an
expansion in $\als$:
\be
F(r)=C_F{\al_{qq}(1/r) \over r^2}={C_F \over
r^2}\als\left\{1+a_{qq}^{(1)}\als+\cdots\right\} 
\,.
\ee
In practice, the force is introduced in Eq. (\ref{force}) order by
order in $\als$. If the expansion parameter is $\als(\nu)$, the
potential is reconstructed order by order in $\als(\nu)$ and we are
exactly in one of the situations considered in this and the previous
sections (Fig. \ref{potOSlattnu}). If the expansion parameter in the
force is $\als(1/r)$, as it was made in Refs. \cite{Sumino,DESY}, we
are in a new situation. The introduction of $\als(1/r)$ in the
integral in Eq. (\ref{force}) produces and infinite series in
$\als(1/r)$ and $\als(1/r_0)$. At this respect, it is not clear what
the expansion parameter is and the systematics of this procedure. In
any case, here, we are just interested to see whether the renormalon
cancellation is achieved for these calculations.  At this respect, the
point is that the expansion parameter in the potential is not
$\als(r)$. From the computational point of view, all the powers in
$\als(1/r)$ and $\als(1/r_0)$ are considered to be of the same
order. If we consider the leading order, one loop, approximation we
have $\al_{qq}=\als$ and $\beta(\al_{qq})=-\beta_0\als/(2\pi)$ and we
obtain
\be
\int_{r_0}^rdr'F(r')=-C_F{\als(1/r) \over r}
\left(
\sum_{n=0}^{\infty}n!\left[{\beta_0 \als(1/r) \over 2\pi}\right]^n
-
\sum_{n=0}^{\infty}n!\left[{\beta_0 \als(1/r_0) \over 2\pi}\right]^n
\right)
\,.
\ee
Both terms have the renormalon (in the large $\beta_0$ approximation)
so that they cancel each other. We expect that higher orders
computations will reconstruct the QCD renormalon but yet the
renormalon cancellation should remain at any order since the same
(partial) piece of the renormalon appears for
$\int^rdr'F^{n-loops}(r')$ than for
$\int^{r_0}dr'F^{n-loops}(r')$. This explains within a
renormalon-dominance approach the convergence of computations using
this framework even if no reference to renormalon was made for these
calculations.

\medskip

\begin{figure}[h]
\makebox[1.0cm]{\phantom b}
\put(-7,137){$r_0(V_{\RS}(r)-V_{\RS}(r')+E_{latt.}(r'))$}
\put(228,137){$r_0(V_{\RS}(r)-V_{\RS}(r')+E_{latt.}(r'))$}
\put(-30,10){\epsfxsize=8truecm \epsfbox{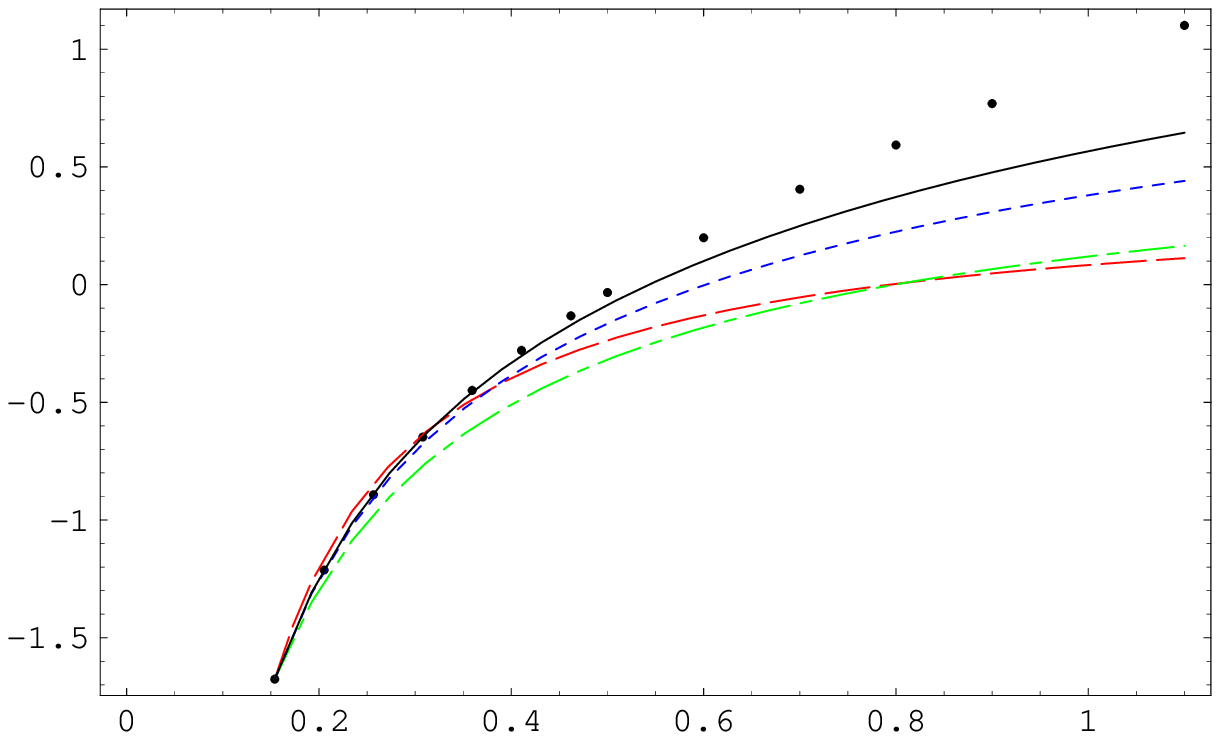}}
\put(210,10){\epsfxsize=8truecm \epsfbox{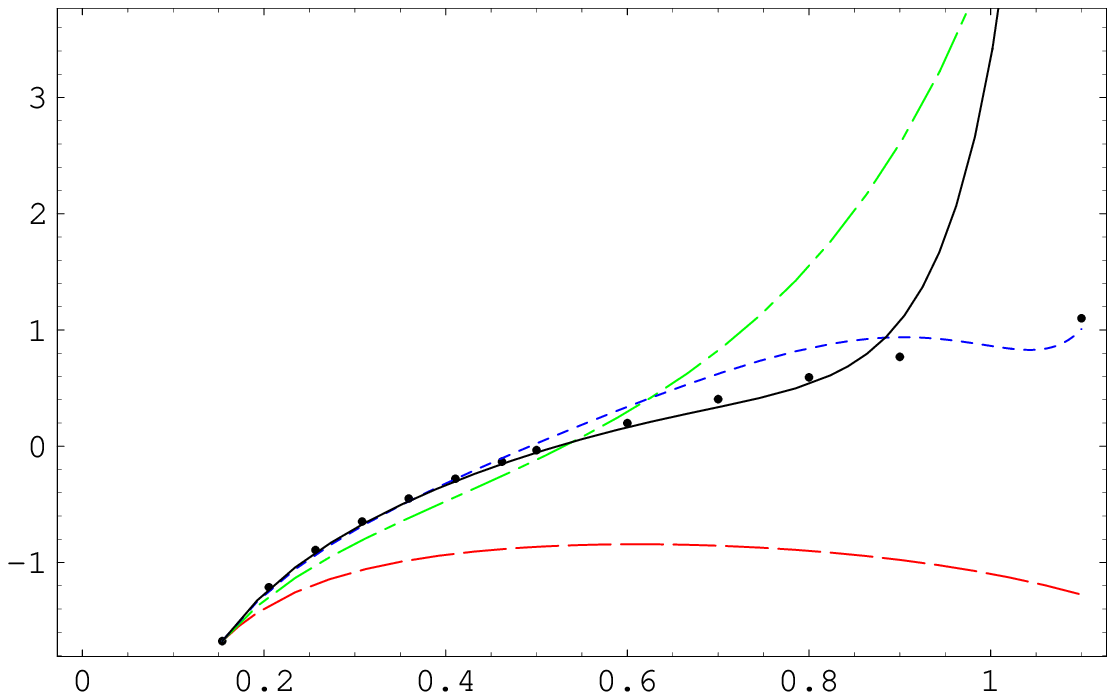}}
\put(100,1){$r/r_0$}
\put(300,1){$r/r_0$}
\put(-15,1){$a)$}
\put(220,1){$b)$}
\caption {{\it Figs. a) and b) correspond to Figs. \ref{potOSlattnu}
and \ref{potRSlattr} respectively up to the value $r=1.1r_0$.}}
\label{combined}
\vspace{1mm}
\end{figure}

It is worth stressing that within this renormalon-based scheme
calculation, we can now account for the logs correctly by setting
$\nu=1/r$ (see Fig. \ref{potRSlattr}).  On the other hand, we
introduce some errors when fixing $\nu=1/r$, since we are only able to
perform an approximate computation of the renormalon, which do not
exist in the calculation of the slope with $\nu=constant$. The two
things compete with each other in the final accuracy of the result
(compare Figs. \ref{potOSlattnu} and \ref{potRSlattr}). Indeed, once
the renormalon cancellation has been achieved, the resummation of logs
does not appear to be very important. At least, if we do not choose
our scale very far from the typical scales at study. If we go to
scales smaller than $2.5\,r_0^{-1}$ , the calculation with $\nu=1/r$
appears to give better results (compatible
with lattice, although with larger errors, up to amazingly long
distances ($\sim 0.8\, r_0$)) than the calculation with fixed $\nu$
(see Fig. \ref{combined}). This may signal that log resummation
becomes important for these scales (note that we have chosen $\nu=1/0.15399
\, r_0^{-1}$ as the starting point of the log evolution, for a 
smaller value of $\nu$ an improvement may be obtained). It should be studied
further whether perturbation theory can indeed describe lattice data
in the regime $0.5-0.8 \, r_0$, since this would support the claims of
Ref. \cite{Sumino} in this direction.

\medskip

The main result of this section is to show that all the different
results found in the previous and this section as well as in the
literature about the convergence (or not convergence) of the
perturbative series can be explained by the presence of the
renormalon. We have also shown that, with a renormalon-based scheme,
agreement with lattice is obtained. The point is that we now know the
constant to be subtracted and its dependence in $\als$, therefore, we
can set $\als$ at the same scale and obtain the cancellation for any
$r$. In this way, we obtain the log resummation and the renormalon
cancellation. One potential problem is that other logs ($\ln(r\nu_f)$)
appear (in the physical, finite mass, case this problem does not
really show up because one usually sets $\nu_f \siml 1/r$ being $r$
the typical size of the system).

\section{Bounds on short-distance non-perturbative potentials}

\begin{figure}[h]
\hspace{-0.1in}
\epsfxsize=5in
\centerline{
\put(50,190){$r_0(V_{\RS}(r)-V_{\RS}(r')+E_{latt.}(r'))$}
\epsffile{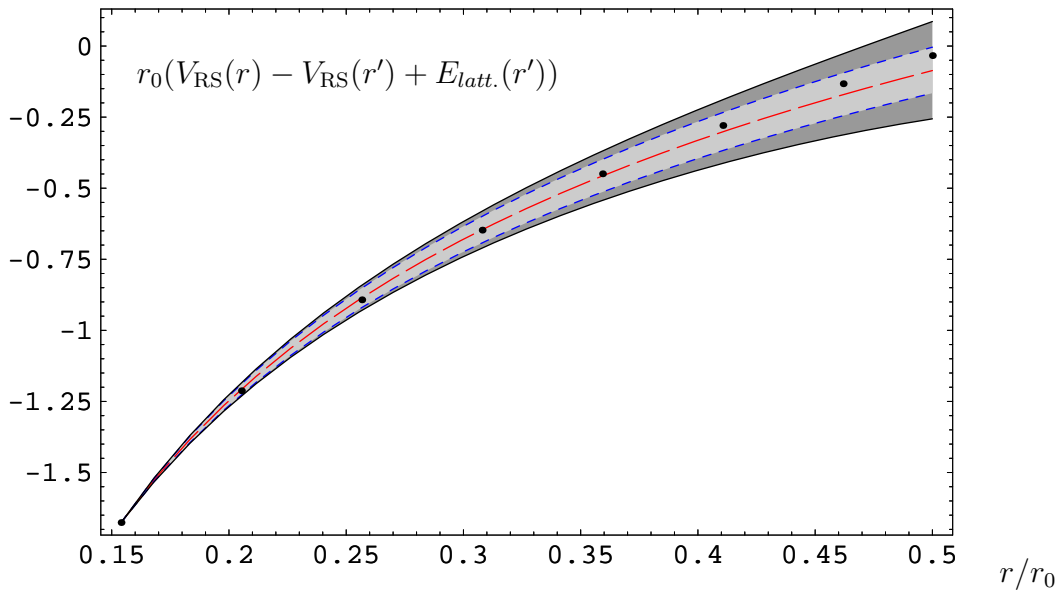}
\put(15,1){$r/r_0$}
}
\caption {{\it Plot of $r_0(V_{\RS}(r)-V_{\RS}(r')+E_{latt.}(r'))$
versus $r$ at three loops (estimate) plus the leading single ultrasoft
log (dashed line) compared with the lattice simulations \cite{NS}
$E_{latt.}(r)$. For the scale of $\als(\nu)$, we set
$\nu=1/0.15399\,r_0^{-1}$. $\nu_{us}=2.5\,r_0^{-1}$ and
$r'=0.15399\,r_0$. The inner and outer band are meant to estimate the
errors in $\Lambda_{\MS}$ and $c_0$. For further details see the main
text.}}
\label{errors}
\end{figure}

We now study possible non-perturbative effects in the static
potential. The first thing to notice is that any non-perturbative
effects should be small and compatible with zero since perturbation
theory is able to explain lattice data within errors. We can make this
statement more quantitative by using the lattice data obtained in
Ref. \cite{NS} where the continuum limit has been reached. For these
lattice points the systematic and statistic errors are very small
(smaller than the size of the points). Therefore, the main sources of
uncertainty of our (perturbative) evaluation come from the uncertainty
in the value of $\Lambda_{\MS}$ ($\pm 0.48\,r_0^{-1}$) obtained from
the lattice \cite{Lambda} and from the uncertainty in higher orders in
perturbation theory. We show our results in
Fig. \ref{errors}.\footnote{We have performed the estimation of the
errors with the results of Fig. \ref{potOSlattnu} ($\nu=constant$),
since they do not introduce errors due to the evaluation of the
renormalon. Nevertheless, a similar conclusion had been achieved if
the analysis had been done with the results of Fig. \ref{potRSlattr}
($\nu=1/r$).} The inner band reflects the uncertainty in
$\Lambda_{\MS}$ whereas the outer band is meant to estimate the
uncertainty due to higher orders in perturbation theory by allowing
$c_0$ to change by $\pm 146$. This is half its value according to the
estimate in Table 1. This may seem a conservative variation if we take
into account that the difference with the estimate obtained using
Pade-approximats \cite{CE} is $\delta c_0 \sim 20$.

\subsection{Bounds on a short-distance quadratic potential}
\label{quadpot}

The non-perturbative effects should be encoded within
Eq. (\ref{energyUS}). In general, this equation is a convolution of
two scales: $\lQ$ and $\als/r$. The explicit functionality is unknown
since it involves the knowledge of the details of the non-perturbative
dynamics. In the formal limit $1/r \gg \lQ \gg \als/r$, it is possible
to know the scaling of the potential in $\lQ$ and $r$ based on
dimensional arguments \cite{Balitsky}:
\begin{equation}
\delta E_{\rm US}(r,\nu_{us}) \simeq {T_F \over 3 N_c} {\bf r}^2 \int_0^\infty \!\! dt 
\langle g{\bf E}^a(t) 
\phi(t,0)^{\rm adj}_{ab} g{\bf E}^b(0) \rangle(\nu_{us}).
\label{energyUSr2}
\end{equation}

This $r^2$ behavior for the non-perturbative potential is expected on
the basis of the OPE in that limit. 

The fact that we are able to put bounds on our perturbative
evaluation of the potential and that the errors of the lattice
simulations are very small allows us to put bounds on the
non-perturbative effects. This is of utmost importance since these
non-perturbative effects produce the largest errors in the
determination of the bottom mass using the $\Upsilon(1S)$
mass. Unfortunately, even though lattice simulations of the static
potential are very precise, we can not obtain strong constraints on
the non-perturbative effects. The main source of error is due to the 
uncertainty in $\Lambda_{\MS}$. Assuming a non-perturbative potential
$V\sim Ar^2$, we obtain 
\be
|A| \siml 0.6\,r_0^{-3}
\ee
from the variation of $\Lambda_{\MS}$. If a potential
$0.6\,r_0^{-3}r^2$ were introduced as a correction in the binding
energy of the $\Upsilon(1S)$, it may induce a correction of $\sim
100\,{\rm MeV}$ in the mass of the $\Upsilon(1S)$ (however, let us
also stress that lattice simulations are also compatible with $A=0$).
Nevertheless, if lattice determinations of $\Lambda_{\MS}$ could be
improved, they would provide one of the most promising ways to control
the errors in the determination of the bottom $\MS$ mass using the
$\Upsilon(1S)$ mass. Let us also note that one can turn the problem
around. If one assumes the non-perturbative effects to be small one
can improve the lattice determination of $\Lambda_{\MS}$ from the
static potential data.

\subsection{Bounds on a short-distance linear potential}

\begin{figure}[h]
\hspace{-0.1in}
\epsfxsize=5in
\centerline{
\put(40,190){$r_0(V_{\RS}(r)-V_{\RS}(r')+E_{latt.}(r')+\sigma (r-r'))$}
\epsffile{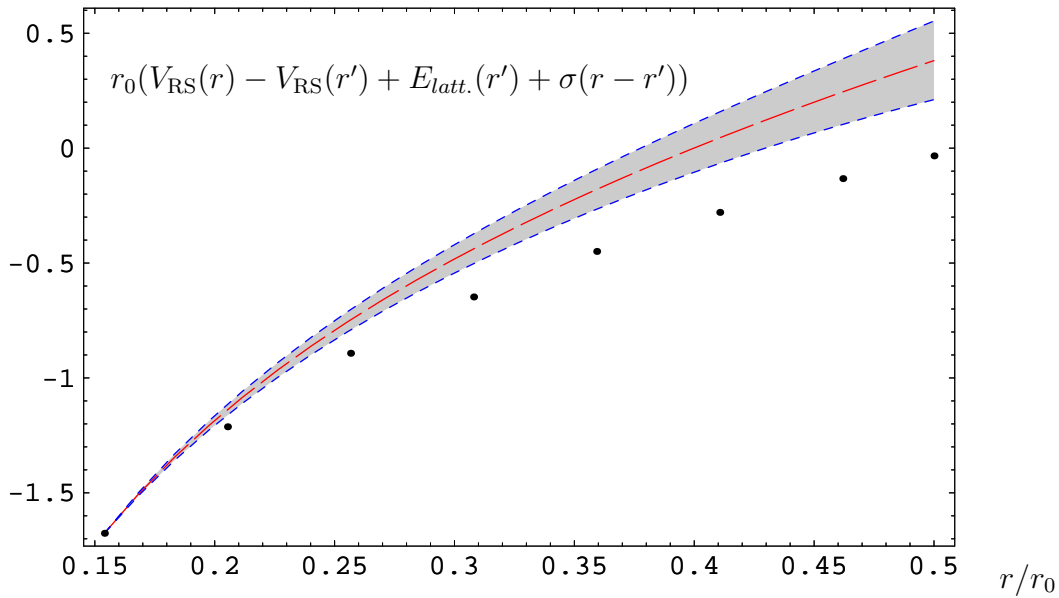}
\put(15,1){$r/r_0$}
}
\caption {{\it Plot of $r_0(V_{\RS}(r)-V_{\RS}(r')+E_{latt.}(r')+\sigma
(r-r'))$ versus $r$ at three loops (estimate) plus the leading single ultrasoft
log compared with the lattice simulations \cite{NS}
$E_{latt.}(r)$. For the scale of $\als(\nu)$, we set
$\nu=1/0.15399\,r_0^{-1}$. $\nu_{us}=2.5\,r_0^{-1}$ and
$r'=0.15399\,r_0$. The dashed band is meant to estimate the
combined error due $\Lambda_{\MS}$ and $c_0$ (see Fig. \ref{errors}).}}
\label{linearpot}
\end{figure}

The usual confining potential, $\delta V =\sigma r$, goes with an
slope $\sigma=0.21 {\rm GeV}^2$. In lattice units we take:
$\sigma=1.35\, r_0^{-2}$. Can we discriminate such potential at short
distances with the available lattice data? The answer is yes as we can
see from Fig. \ref{linearpot}.\footnote{We have used the results from
Fig. \ref{potOSlattnu}, since they do not introduce errors due to the
evaluation of the renormalon. Nevertheless, the same conclusion had
been achieved with the results from Fig. \ref{potRSlattr}.} The
introduction of a linear potential at short distances with such slope
is not consistent with lattice simulations. This is even so after the
errors considered in Fig. \ref{errors} have been included. This should
not come as a surprise since this linear potential appears as an
effect of long distance (not short distance) and it is not expected to
appear from Eq. (\ref{energyUS}).  Therefore, it follows that the use
of the Cornell potential (with the perturbative static potential
emanated from QCD instead of a pure $1/r$ potential times a fitted
constant) as a phenomenological fit of the static potential introduces
systematic errors if the typical inverse Bohr radius scale of the
heavy quarkonium system to study lye in the short distance regime as
it is the case, for instance, for the $\Upsilon(1S)$.
\medskip

\medskip

On the other hand, recently, there have been claims about the possible
existence of a linear potential at short distances
\cite{GPZ,S}. Expected it to be of different physical origin than the
long distance linear potential, it may have a different slope than the
(long distance) static potential discussed above. It would be very
important to discriminate its existence, since such behavior at short
distances is at odds with the OPE.  This is indeed possible, since the
short-distance linear potential expected in Ref. \cite{GPZ,S} have an
slope of the order of magnitude of the long-distance confining
potential that we have already ruled out above. Therefore, we can
conclude that no linear potential exists at short distance (with the
present estimates for its slope).

\section{Conclusions}

In this paper, we have shown (Figs. \ref{potOSlattnu} and
\ref{potRSlattr}) that perturbation theory can explain lattice
simulations of the static potential up to scales of order $\sim
2\,r_0^{-1}$ once the renormalon has been taken into account without
the need of non-perturbative effects.  Quite remarkable, lattice
simulations are precise enough to see the scaling log behavior
predicted by asymptotic freedom and not just the pure $1/r$ behavior
of the Coulomb potential.

We have shown that the different results one may find in the
literature as well as in sections 2 and 3 about the convergence (or
not) of the perturbative series to the lattice result can be explained
by the presence of the renormalon. In particular, we can explain why a
calculation (Fig. \ref{potOSlattnu}) within perturbation theory can
agree with lattice data, if the perturbative expansion is made with
$\als(\nu)$ with $\nu$ constant, without an explicit reference to the
renormalon, but not if the expansion is made with $\als(1/r)$
(Fig. \ref{potOSlattr}). This explanation is given in a
renormalon-based approach in a unified way. Moreover, our method
allows to set $\nu=1/r$ and, therefore, resum the $\ln(r\nu)$ logs
besides of cancelling the renormalon (Fig. \ref{potRSlattr}).

We have tried to find bounds on possible non-perturbative potentials
at short distances.  As a matter of principle, large non-perturbative
effects are ruled out with the above analysis. In particular, they are
compatible with zero (Fig. \ref{errors}).  We have shown
(Fig. \ref{linearpot}) that lattice data disfavours a linear confining
potential at short distances with the usual slope
($\sigma=1.35\,r_0^{-2}$) assigned to the confining potential at long
distances. For a possible non-perturbative $r^2$ potential, the main
source of error comes from the one associated to the determination of
$\Lambda_{\MS}$ in the lattice (the error due to higher order in
perturbation theory appears to be less important). Therefore, we can
not put very precise constraints on it in order to reduce the
non-perturbative error on the determinations of the bottom $\MS$ mass
from the $\Upsilon(1S)$ mass.  Nevertheless, lattice simulations
remain as one of the most promising possibilities to constraint the
size of the non-perturbative effects at short distances.

\medskip

{\bf Acknowledgments}\\
We thank G.S. Bali for sharing his lattice data with us.



\begin{thebibliography}{99}


\bibitem{latticeshort1} G.S.~Bali and K.~Schilling,
Phys.\ Rev.\  {\bf D47}  (1993) 661; 
G.S.~Bali and K.~Schilling,
Int.\ J.\ Mod.\ Phys.\  {\bf C4} (1993) 1167; 
G.S.~Bali, K.~Schilling and A.~Wachter,
Phys.\ Rev.\  {\bf D56} (1997) 2566.

\bibitem{NS} S. Necco and R. Sommer, Nucl. Phys. {\bf B622}, 328 (2002). 

\bibitem{latticeshort2} K.J. Juge, J. Kuti and C.J. Morningstar,
Nucl. Phys. (Proc. Suppl.) {\bf 83}, 503 (2000). 
 
\bibitem{FSP} W. Fischler, Nucl. Phys. {\bf B129}, 157 (1977);
Y. Schr\"oder, Phys. Lett. {\bf B447}, 321 (1999); B.A. Kniehl,
A.A. Penin, V.A. Smirnov and M. Steinhauser, Phys. Rev. {\bf D65},
091503 (2002); M. Peter, Phys. Rev. Lett. {\bf 78}, 602 (1997).

\bibitem{short} N. Brambilla, A. Pineda, J. Soto and A. Vairo,
     Phys. Rev. {\bf D60} (1999) 091502. 

\bibitem{KP1} B.A. Kniehl and A.A. Penin, Nucl. Phys. {\bf B563}, 200
  (1999).

\bibitem{RG} A. Pineda and J. Soto, Phys. Lett. {\bf B495}, 323 (2000).

\bibitem{HMS} A.H. Hoang, A.V. Manohar and I.W. Stewart,
Phys. Rev. {\bf D64}, 014033 (2001).

\bibitem{PS} A.A. Penin and M. Steinhauser, private communication. 

\bibitem{Aglietti} U. Aglietti and Z. Ligeti, Phys. Lett. {\bf
    B364}, 75 (1995).

\bibitem{thesis} A. Pineda, PhD. thesis, U. Barcelona, January
(1998); A.H. Hoang, M.C. Smith, T. Stelzer and S. Willenbrock,
Phys. Rev. {\bf D59}, 114014 (1999); 
M. Beneke, Phys. Lett. {\bf B434}, 115 (1998).

\bibitem{Bali} G.S. Bali, Phys. Lett. {\bf B460}, 170 (1999).

\bibitem{GPZ} F.G. Gubarev, M.I. Polikarpov and V.I. Zakharov,
Mod. Phys. Lett. {\bf A14}, 2039 (1999).  

\bibitem{S} Yu.A. Simonov, JETP Lett. {\bf 69}, 505 (1999).

\bibitem{BSV} N. Brambilla, Y. Sumino and A. Vairo, Phys. Lett. {\bf
B513}, 381 (2001); Phys. Rev. {\bf D65}, 034001 (2002). 

\bibitem{Sumino} Y. Sumino, Phys. Rev. {\bf D65}, 054003 (2002); 
S. Recksiegel and Y. Sumino, Phys. Rev. {\bf D65}, 054018 (2002).

\bibitem{HLM} A. Hoang, Z. Ligeti and A.V. Manohar, Phys. Rev. Lett. {\bf 82}, 
277 (1999).

\bibitem{DESY} S. Necco and R. Sommer, Phys. Lett. {\bf B523}, 135 (2001).

\bibitem{BB} P. Boyle and G.S. Bali, Nucl. Phys. {\bf B}
(Proc. Suppl.) {\bf 106}, 811 (2002).

\bibitem{RS} A. Pineda, {\bf JHEP0106}, 022 (2001).

\bibitem{pNRQCD} A. Pineda and J. Soto, Nucl. Phys. {\bf B}
(Proc. Suppl.) {\bf 64}, 428 (1998); N. Brambilla, A. Pineda, J. Soto
and A. Vairo, Nucl. Phys. {\bf B566}, 275 (2000).

\bibitem{CE} F.A. Chishtie and V. Elias, Phys. Lett. {\bf B521}, 434 (2001).

\bibitem{Lambda} ALPHA collaboration: S. Capitani, M. L\"uscher,
R. Sommer and H. Wittig, Nucl. Phys. {\bf B544}, 669 (1999).

\bibitem{Beneke2} M. Beneke, Phys. Lett. {\bf B344}, 341 (1995). 

\bibitem{renormalons} M. Beneke, Phys. Rep. {\bf 317}, 1 (1999).

\bibitem{Lee} T. Lee, Phys. Rev. {\bf D56}, 1091 (1997);
Phys. Lett. {\bf B462}, 1 (1999).

\bibitem{Balitsky} I.I. Balitsky, Nucl. Phys. {\bf B254}, 166 (1985).

\end{thebibliography}
\end{document}